\documentclass[aps,preprint,prb]{revtex4}
\usepackage{amsmath}
\usepackage{graphicx}

\input{tcilatex}
\begin{document}

\title{Density Functional Theory of Inhomogeneous Liquids: II. A Fundamental
Measure Approach}
\author{James F. Lutsko}
\affiliation{Physics Department CP 231, Universit\'{e} Libre de Bruxelles, Blvd. du
Triomphe, 1050 Brussels, Belgium}
\email{jlutsko@ulb.ac.be}
\homepage{http://www.lutsko.com}

\begin{abstract}
Previously, it has been shown that the direct correlation function for a
Lennard-Jones fluid could be modeled by a sum of that for hard-spheres , a
mean-field tail and a simple linear correction in the core region
constructed so as to reproduce the (known) bulk equation of state of the
fluid(Lutsko, JCP 127, 054701 (2007)). Here, this model is combined with
ideas from Fundamental Measure Theory to construct a density functional
theory for the free energy. The theory is shown to accurately describe a
range of inhomogeneous conditions including the liquid-vapor interface, the
fluid in contact with a hard wall and a fluid confined in a slit pore. The
theory gives quantitatively accurate predictions for the surface tension,
including its dependence on the potential cutoff. It also obeys two
important exact conditions: that relating the direct correlation function to
the functional derivative of the free energy with respect to density, and
the wall theorem.
\end{abstract}

\date{\today }
\maketitle

\section{Introduction}

The insights of Van der Waals (VDW)\cite{VDW1,VDW2} still underlie much of
the work on nonuniform fluids. One such insight is that the free energy of a
fluid can be separated into two contributions:\ the first due to the
short-ranged repulsion of all interaction models, the second due to a
long-ranged attraction. The former is generally treated using an effect
hard-sphere contribution and the latter by a mean-field approximation. This
model is attractive because of its simplicity and the prospect that it can
be extended to complex systems such as molecular fluids, anisotropic
interactions etc. However, while it can generally give reasonable
qualitative predictions, such a simple model is seldom quantitatively
accurate. It would therefore be highly desirable to modify the basic VDW\
model so as to give quantitatively accurate predictions if this can be done
without compromising the basic simplicity of the model. The goal of this
paper is to describe such a model and to show by application to the
liquid-vapor interface and to fluids in confined geometries that it does
indeed satisfy the twin requirements of accuracy and simplicity.

In the language of density functional theory, the VDW model can be described
as a particular approximation to the direct correlation function (DCF) of the
bulk fluid:\ namely that it is the combination of a hard-sphere DCF and a
mean-field tail. In a previous paper\cite{Lutsko_JCP_2007} it was noted that
the failure of this model for the DCF lies primarily within the core region
and that a good first order correction could be obtained by adding to the
hard-sphere DCF a simple linear correction with coefficients adjusted to
give continuity of the DCF and to give the correct thermodynamics. (The bulk
thermodynamics are assumed to be known from e.g. thermodynamic perturbation
theory or liquid state theory.) This is a satisfying approach as one of the
motivations behind DFT (particularly the effective-liquid approaches) has
been the idea that the DCF is a relatively simple function which could be
easily approximated, in contrast to, say, the pair-distribution function
which is highly structured. However, having a model DCF for the \emph{bulk}
fluid is only a first step towards describing non-uniform fluids: this model
must somehow be used to construct a free-energy functional for nonuniform
systems. If the DCF for an arbitrary nonuniform system were known, the free
energy could be obtained immediately since the DCF is the second functional
derivative of the free energy with respect to density. Thus one approach is
to guess a generalization of the known DCF for a uniform system. In fact,
since such functionals are known for the hard-sphere system and since the
mean-field tail is independent of density, the generalization need only be
guessed for the core-correction which limits the problem. In ref. \cite%
{Lutsko_JCP_2007}, the usual ideas from DFT, involving the introduction of
local densities into bulk expressions, were used to construct such a
functional with mixed success: while the model gave reasonable predictions
for the surface tension of the liquid-vapor interface, it suffered from some
arbitrary elements due to the fact that the DCF and the proposed free energy
functionals were internally inconsistent. That work therefore served to
demonstrate the utility of trying to correct the DCF but did not fully
address the problem of constructing a satisfactory free energy functional.

In the case of hard-spheres, the problem of constructing a widely useful
free energy functional has been solved in recent years by the development of
Fundamental Measure Theory (FMT)\cite{Rosenfeld1,Rosenfeld2}. Rosenfeld
originally proposed FMT as a generalization of ideas from scaled particle
theory\cite{Rosenfeld1}, but for present purposes, one of the most
interesting derivations of FMT is that of Kierlik and Rosinberg\cite{Kierlik}%
. They get a theory essentially equivalent to Rosenfeld's by starting with
the same general ansatz for the free energy functional (which, they point
out, is the obvious generalization of the exact result known for
one-dimensional systems) and requiring that the second functional derivative
of this ansatz with respect to density give the Percus-Yevick DCF in the
uniform limit. Thus, in this context, FMT can be seen as the result of a
constructive exercise in which one starts with a known DCF, a particular
ansatz for the free energy functional (which is exact in one dimension) and
enforces the exact relation between the DCF and the free energy functional.
Here, I propose that the the same constructive procedure be used to
incorporate the core correction giving a fully consistent relation between
the DCF and the free energy functional. Since the result is a
straightforward modification of the hard-sphere contribution, while the
mean-field tail is unchanged, the resulting model also satisfies the
requirement for simplicity. This model will be referred to as the Modified
Core VDW\ or MC-VDW\ free energy.

The MC-VDW model requires as input the equation of state of the bulk fluid.
This is in keeping with the view adopted in ref. \cite{Lutsko_JCP_2007} that
the main purpose of a model DFT is not to compute the properties of bulk
fluids, which can be done very accurately using thermodynamic perturbation
theory or liquid state theory, but to be used to calculate the properties of 
\emph{inhomogeneous} fluids. This requirement is relatively mild compared to
older DFTs that required knowledge of the direct correlation function of the
bulk fluid as input. In the following, the Lennard-Jones system will be
studied using both a perturabtive equation of state and an empirical
equation of state.

In the next section, the details of the MC-VDW model are given. The third
section consists of a comparison of the predictions of the model to data
from computer simulations. The first comparison is the surface tension and
density profile at the planar interface between coexisting liquid and vapor
phases. It is shown that not only is the surface tension as a function of
temperature accurately predicted, but so is the variation of the surface
tension with the range of the potential. The second comparison is that of
the structure of the fluid at a hard wall. It is noted that the theory
satisfies the exact sum rule known as the Wall Theorem relating the density
at the wall to the pressure far from the wall. The density as a function of
distance to the wall is compared to the simulation results for several
different temperatures. The final comparison is that of the structure of the
fluid in slit pores. In all cases, the theory is found to be in good
quantitative agreement with the simulations. The paper ends with a
discussion of the results and of possible future developments.

\section{Theory}

Given a collection of atoms interacting via a spherically symmetric pair
potential, $v(r)$ at fixed temperature $T$, chemical potential $\mu $ and
external one-body field $V_{ext}\left( \mathbf{r}\right) $, the equilibrium
density distribution $\rho \left( \mathbf{r}\right) $ is obtained by
minimizing the functional 
\begin{equation}
\beta \Omega \left[ \rho \right] =\int d\mathbf{r}\;\left[ \rho \left( 
\mathbf{r}\right) \ln \rho \left( \mathbf{r}\right) -\rho \left( \mathbf{r}%
\right) +\beta f^{\times }\left( \mathbf{r};\left[ \rho \right] \right)
-\beta \mu \rho \left( \mathbf{r}\right) +\rho \left( \mathbf{r}\right)
\beta V_{ext}\left( \mathbf{r}\right) \right]
\end{equation}%
where $\beta =1/k_{B}T$ and $k_{B}$ is Boltzmann's constant\cite%
{MerminDFT,HansenMcdonald}. The value of the functional at its minimum is
the grand potential for the system. The only unknown term here is the excess
free energy density, $f^{\times }\left( \mathbf{r};\left[ \rho \right]
\right) $. It is related to the DCF for a nonuniform system via%
\begin{equation}  \label{exactDCF}
c\left( \mathbf{r}_{1},\mathbf{r}_{2};\left[ \rho \right] \right) =-\frac{%
\delta ^{2}\beta F^{x}}{\delta \rho \left( \mathbf{r}_{1}\right) \delta \rho
\left( \mathbf{r}_{2}\right) }\mathbf{,}
\end{equation}%
where%
\begin{equation}
\beta F^{x}=\int \beta f^{\times }\left( \mathbf{r};\left[ \rho \right]
\right) d\mathbf{r.}
\end{equation}%
In ref.\cite{Lutsko_JCP_2007}, it was shown by a direct comparison to
computer simulation that the DCF for the bulk fluid can be adequately
approximated by a model of the form%
\begin{equation}
c\left( r_{12};\left[ \rho \right] \right) =c_{HS}\left( r_{12};d,\left[
\rho \right] \right) +\left( a_{0}+a_{1}\frac{r_{12}}{d}\right) \Theta
\left( d-r_{12}\right) -\beta w\left( r_{12}\right) ,  \label{DCF}
\end{equation}%
where the first term on the right is the hard-sphere DCF, $d$ is the
Barker-Henderson effective hard-sphere diameter, the constants $a_{0}$ and $%
a_{1}$, which are functions of both density and temperature, are determined
by requiring that this DCF gives the correct bulk free energy and that the
DCF be continuous at the hard-sphere boundary $r_{12}=d$ (see ref. \cite%
{Lutsko_JCP_2007} and Appendix \ref{Model} for explicit expressions). The
terms involving $a_0$ and $a_1$ are referred to below as the "core
correction". There are several reasonable choices for the tail function $%
\beta w\left( r_{12}\right) $ but here I only consider the simplest choice, $%
w\left( r_{12}\right) =\Theta \left( r_{12}-d\right) v\left( r_{12}\right) $%
. The idea is to use this model DCF for the homogeneous fluid together with
the expression relating the DCF to the excess free energy functional to
guide the construction of a functional that can be used for inhomogeneous
fluids.

In the simplest form of FMT\cite{Rosenfeld1}, the excess free energy for
hard-spheres of diameter $d$ is given by%
\begin{equation}
f_{HS}^{\times }\left( \mathbf{r};\left[ \rho \right] \right) =s\left( 
\mathbf{r}\right) \Phi _{1}\left( \eta \left( \mathbf{r}\right) \right)
+\Phi _{2}\left( \eta \left( \mathbf{r}\right) \right) \left( s^{2}\left( 
\mathbf{r}\right) -v^{2}\left( \mathbf{r}\right) \right) +\Phi _{3}\left(
\eta \left( \mathbf{r}\right) \right) s\left( \mathbf{r}\right) \left(
s^{2}\left( \mathbf{r}\right) -3v^{2}\left( \mathbf{r}\right) \right)
\end{equation}%
 where the explicit form of the algebraic functions $\Phi_i(\eta)$ are given in ref. \cite{Rosenfeld1}. The quantities $n_{\alpha }\left( \mathbf{r}\right) =\left( \eta
\left( \mathbf{r}\right) ,s\left( \mathbf{r}\right) ,\mathbf{v}\left( 
\mathbf{r}\right) \right) $ are linear functionals of the density 
\begin{equation}
n_{\alpha }\left( \mathbf{r}\right) =\int w_{\alpha }\left( \left\vert 
\mathbf{r}-\mathbf{r}_{1}\right\vert \right) \rho \left( \mathbf{r}%
_{1}\right) d\mathbf{r}_{1}.
\end{equation}%
The weights $w_{\alpha }\left( r\right) $ are $\Theta(\frac{d}{2}-r)$, $%
\delta(\frac{d}{2}-r)$ and $\frac{\mathbf{r}}{r}\delta(\frac{d}{2}-r)$,
respectively. In the uniform limit, the density becomes a constant $\rho
\left( \mathbf{r}\right) \rightarrow \overline{\rho }$ and the quantity $%
\eta \left( \mathbf{r}\right) \rightarrow \pi \overline{\rho }d^{3}/6$,
which is the usual expression for the packing fraction. The other
functionals become $s\left( \mathbf{r}\right) \rightarrow \pi \overline{\rho 
}d^{2}$and $\mathbf{v}\left( \mathbf{r}\right) \rightarrow 0$. It is useful
to introduce dimensionless quantities via%
\begin{eqnarray}
\Phi _{1}\left( \eta \right) &=&\frac{1}{\pi d^{2}}h_{1}\left( \eta \right)
\\
\Phi _{2}\left( \eta \right) &=&\frac{1}{\pi d}h_{2}\left( \eta \right) 
\notag \\
\Phi _{3}\left( \eta \right) &=&\frac{1}{\pi }h_{3}\left( \eta \right) 
\notag
\end{eqnarray}%
It is then straightforward to show that 
\begin{equation}
\lim_{\rho \left( \mathbf{r}\right) \rightarrow \overline{\rho }}\frac{%
\delta ^{2}\beta F^{x}_{HS}}{\delta \rho \left( \mathbf{r}_{1}\right) \delta
\rho \left( \mathbf{r}_{2}\right) }=2\Theta \left( d-r_{12}\right) \left[ 
\begin{array}{c}
\left( 6\eta h_{1}^{\prime \prime }\left( \eta \right) +\left( 6\eta \right)
^{2}h_{2}^{\prime \prime }\left( \eta \right) +\left( 6\eta \right)
^{3}h_{3}^{\prime \prime }\left( \eta \right) \right) \left( \frac{1}{12}-%
\frac{1}{8}x+\frac{1}{24}x^{3}\right) \\ 
+\left( 2h_{1}^{\prime }\left( \eta \right) +4\left( 6\eta \right)
h_{2}^{\prime }\left( \eta \right) +6\left( 6\eta \right) ^{2}h_{3}^{\prime
}\left( \eta \right) \right) \left( \frac{1}{4}-\frac{1}{4}x\right) \\ 
+\left( 2h_{2}\left( \eta \right) +6\left( 6\eta \right) h_{3}\left( \eta
\right) \right) \left( \frac{1}{2}x\right)%
\end{array}%
\right]  \label{riv}
\end{equation}%
where $x=r_{12}/d$. For hard spheres, one way to determine the functions $%
h_{i}\left( \eta \right) $ is to compare this expression to that for the
Percus-Yevick DCF.

Here, it is proposed to use the Modified-Core VDW free energy functional 
\begin{equation}
\beta F^{x}=\int \left( \beta f_{HS}^{\times }\left( \mathbf{r};\left[ \rho %
\right] \right) +\beta f_{core}^{\times }\left( \mathbf{r};\left[ \rho %
\right] \right) \right) d\mathbf{r+}\frac{1}{2}\int \rho \left( \mathbf{r}%
_{1}\right) \rho \left( \mathbf{r}_{2}\right) \Theta \left( r_{12}-d\right)
v\left( r_{12}\right) d\mathbf{r}_{1}d\mathbf{r}_{2}  \label{MC-VDW}
\end{equation}%
where the core correction is of the FMT form 
\begin{equation}
\beta f_{core}^{\times }\left( \mathbf{r};\left[ \rho \right] \right) =\frac{%
1}{\pi d^{2}}j_{1}\left( \eta \left( \mathbf{r}\right) \right) s\left( 
\mathbf{r}\right) +\frac{1}{\pi d}j_{2}\left( \eta \left( \mathbf{r}\right)
\right) \left( s^{2}\left( \mathbf{r}\right) -v^{2}\left( \mathbf{r}\right)
\right) +\frac{1}{\pi }j_{3}\left( \eta \left( \mathbf{r}\right) \right)
s\left( \mathbf{r}\right) \left( s^{2}\left( \mathbf{r}\right) -3v^{2}\left( 
\mathbf{r}\right) \right) .
\end{equation}%
The functions $j_{i}\left( \eta \right) $ are introduced as analogs of the
h-functions described above. They are determined by requiring that the model
DCF be recovered in the uniform limit. Comparing eq.(\ref{riv}) and the core
correction in eq.(\ref{DCF}) gives 
\begin{eqnarray}
0 &=&j_{1}^{\prime \prime }+\left( 6\eta \right) j_{2}^{\prime \prime
}+\left( 6\eta \right) ^{2}j_{3}^{\prime \prime }  \label{c1} \\
-a_{0}-a_{1} &=&2j_{2}+6\left( 6\eta \right) j_{3}  \notag \\
-a_{0} &=&j_{1}^{\prime }+2\left( 6\eta \right) j_{2}^{\prime }+3\left(
6\eta \right) ^{2}j_{3}^{\prime }  \notag
\end{eqnarray}%
Its worth noting that use of these relations together with the explicit
expressions for the coefficients $a_{0}$ and $a_{1}$ allows one to prove
that these expressions do indeed reproduce the input bulk free energy in the
uniform limit,%
\begin{eqnarray}
\frac{1}{N}\beta F_{core}^{x} & \equiv & \frac{1}{N}\beta F^{x}-\frac{1}{N}%
\beta F_{HS}^{x}-\frac{1}{2}\overline{\rho }\int \Theta \left( r-d\right)
v\left( r\right) dr  \notag \\
& = & j_{1}(\eta)+6\eta j_{2}(\eta)+\left( 6\eta \right) ^{2}j_{3}(\eta),
\label{c2}
\end{eqnarray}%
where $N$ is the number of atoms, $F^{x}$ is the (known) bulk free energy of
the fluid and $F_{HS}^{x}$ is the FMT hard-sphere free energy functional.
The remainder of the derivation is given in Appendix \ref{Model} and only
the final results will be given here. The function $j_{3}(\eta)$ turns out
to be 
\begin{equation}
j_{3}\left( \eta \right) =\frac{1}{36\eta ^{2}}\left[ \frac{1}{2}\left( 
\frac{1}{N}\beta F_{core}^{x}-\rho \frac{\partial }{\partial \rho }\frac{1}{N%
}\beta F_{core}^{x}-\frac{1}{N}\beta F_{core}^{x}\left( 0\right) \right)
+3\eta \chi ^{HS}\left( \eta \right) -3\int_{0}^{\eta }\chi ^{HS}\left( \eta
\right) d\eta \right]  \label{c3}
\end{equation}%
where $\chi^{HS}\left( \eta \right) =\lim_{r\uparrow d}c_{HS}\left( r;\rho
;d\right) $ is the hard-sphere DCF at the core boundary evaluated at the
density corresponding to the packing fraction $\eta $. (The hard-sphere DCF 
is completely determined by the hard-sphere FMT model: in the Rosenfeld model it is just the Percus-Yevick DCF\cite{Rosenfeld}.) Note that despite the
fact that this expression results from the integration of the differential
equations in eq.(\ref{c1}), there are no integration constants. As shown in
Appendix \ref{Model}, any integration constants are forced to be zero by the
requirement that the functions $j_i(\eta)$ be finite at $\eta=0$. While this
is not shown to be strictly necessary, it seems likely that divergences at
zero density could lead to problems in inhomogeneous systems. Once $%
j_{3}(\eta)$ is determined, the remaining functions, $j_{1}(\eta)$ and $%
j_{2}(\eta)$, follow immediately using the second line of eq.(\ref{c1}) and
eq.(\ref{c2}).

It is easy to see that, by construction, this free energy functional is
consistent in the sense that eq.(\ref{exactDCF}), evaluated in the bulk
limit, gives the assumed model DCF, eq.(\ref{DCF}).

\section{Comparison to simulation}

\begin{figure*}[h!tb]
\resizebox{12cm}{!}{\centerline{\includegraphics[angle=-90]{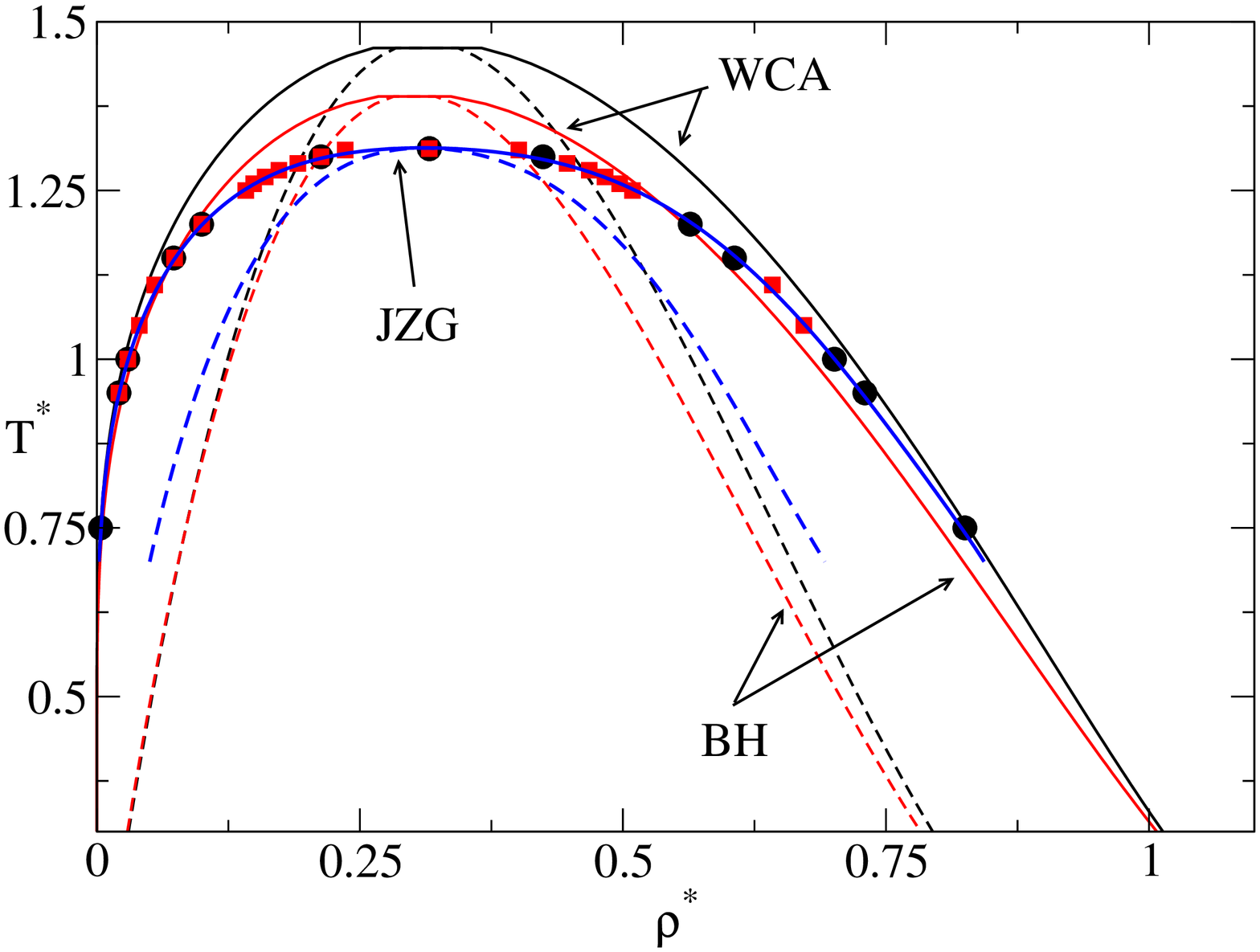}}}
\caption{}
\label{fig1}
\end{figure*}

In this Section, the results of the MC-VDW\ model will be compared to
simulation results for the Lennard-Jones potential,%
\begin{equation}
v_{LJ}\left( r\right) =4\varepsilon \left( \left( \frac{\sigma }{r}\right)
^{12}-\left( \frac{\sigma }{r}\right) ^{6}\right) .
\end{equation}%
Some results will also be given for the truncated and shifted potential,%
\begin{equation}
v_{LJ}\left( r;r_{c}\right) =\left\{ 
\begin{array}{c}
v_{LJ}\left( r\right) -v_{LJ}\left( r_{c}\right) ,\;\;r<r_{c} \\ 
0,\;\;r>r_{c}%
\end{array}%
,\right.
\end{equation}%
which is typically used in Monte Carlo simulations. The only input required
for the model is the equation of state of the bulk fluid. To test the idea
behind the model, apart from other approximations, the very accurate, but
empirical, 33-parameter equation of state of Johnson, Zollweg and Gubbins
(JZG)\cite{JZG} will be used. In order to illustrate the accuracy using more
approximate methods that can be applied to other problems, the results using
the first-order thermodynamic perturbation theory of Barker and Henderson
(BH)\cite{BarkerHend,HansenMcdonald}, and Weeks, Chandler and Anderson (WCA)%
\cite{WCA1,WCA2,WCA3,HansenMcdonald} will also be given. To provide some
context, Fig. \ref{fig1} shows the phase diagram calculated using all three
of these together with simulation data. Clearly, the empirical equation of
state is in close agreement with the simulations whereas the perturbative
theories are reasonable at low temperatures but increasingly inaccurate as
the critical point is approached, as is to be expected.

Details concerning the numerical methods used in minimizing the free energy
functional can be found in ref. \cite{Lutsko_JCP_2007}. In the following,
temperature, $T$, and distance, $z$, will be expressed in reduced units as $%
T^*=T/\epsilon$ and $z^*=z/\sigma$ respectively. A reduced density $%
\rho^*=\rho\sigma^3$ will also be used. The hard-sphere contribution to the
free energy was modeled using the White-Bear FMT functional\cite%
{WhiteBear,tarazona_2002_1} which is somewhat more accurate than the
simplest FMT discussed above. However, for the class of problems considered
here, it probably makes little difference which functional is used.

\subsection{The planar liquid-vapor interface}

\begin{figure*}[h!tb]
\resizebox{12cm}{!}{\centerline{\includegraphics[angle=-90]{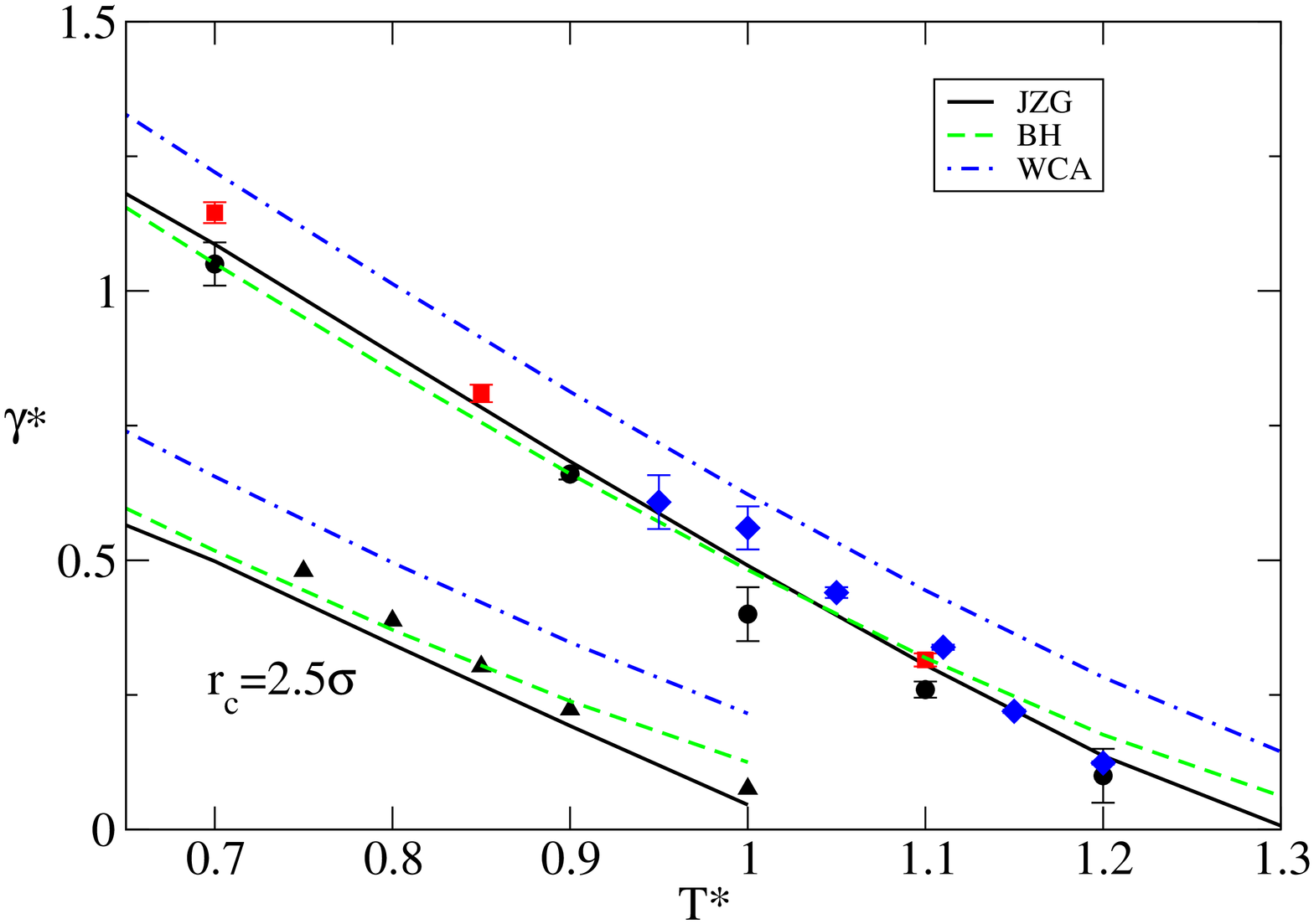}}}
\caption{}
\label{fig2}
\end{figure*}

The planar liquid-vapor interface is determined by minimizing the free
energy functional for a value of the chemical potential corresponding to
liquid-vapor coexistence and only allowing the density to vary in one
direction (the z-direction) and with no external field. If the densities of
the coexisting liquid and vapor are $\rho _{l}$ and $\rho _{v}$,
respectively, then by definition of coexistence, the grand potentials of the
two bulk phases are identical, $\beta \Omega \left( \rho _{l}\right) =\beta
\Omega \left( \rho _{v}\right) $. The excess free energy per unit surface,
the surface tension, is then unambiguously defined as 
\begin{equation}
\gamma =\frac{1}{A}\left( \Omega \left[ \rho \right] -\Omega \left( \rho
_{l}\right) \right)
\end{equation}%
where $A$ is the area of the surface perpendicular to the $z$-axis. The
dimensionless surface tension is $\gamma^*=\sigma^2 \gamma/\epsilon$. Figure
(\ref{fig2}) shows the surface tension as a function of temperature as
calculated from the theory and determined from simulations. Using the
empirical equation of state, the calculated surface tension is consistent
with the data given the scatter in the latter. The perturbative equations of
state given reasonable values although the BH\ perturbative theory is
somewhat superior to the WCA theory. Both are increasingly inaccurate at
higher temperature due to their poor estimation of the critical point, where
the surface tension vanishes.

The figure also shows the calculated surface tension for a cutoff of $%
r_c^*=2.5$ which corresponds to that used in the simulations of Haye and
Bruin\cite{HayeBruin}. Using the empirical equation of state, the theory is
somewhat less accurate in predicting the surface tension than in the case of
the full potential, but the agreement is still reasonable. This is
particularly the case when it is noted that the theory and simulation appear
to extrapolate to slightly different critical temperatures which would
account for most of the discrepancy and which indicates that it originates
in the input equation of state. (Note that the modification of the JZG
equation of state needed to account for the cutoff is not exact and is most
inaccurate for very short cutoffs\cite{JZG}.) The results using the
approximate equations of state are similar to those found with the full
potential: the BH theory gives quantitatively better results, again probably
due to the fact that it gives a better estimate of the critical point.

\begin{figure*}[h!tb]
\resizebox{12cm}{!}{\centerline{\includegraphics[angle=-90]{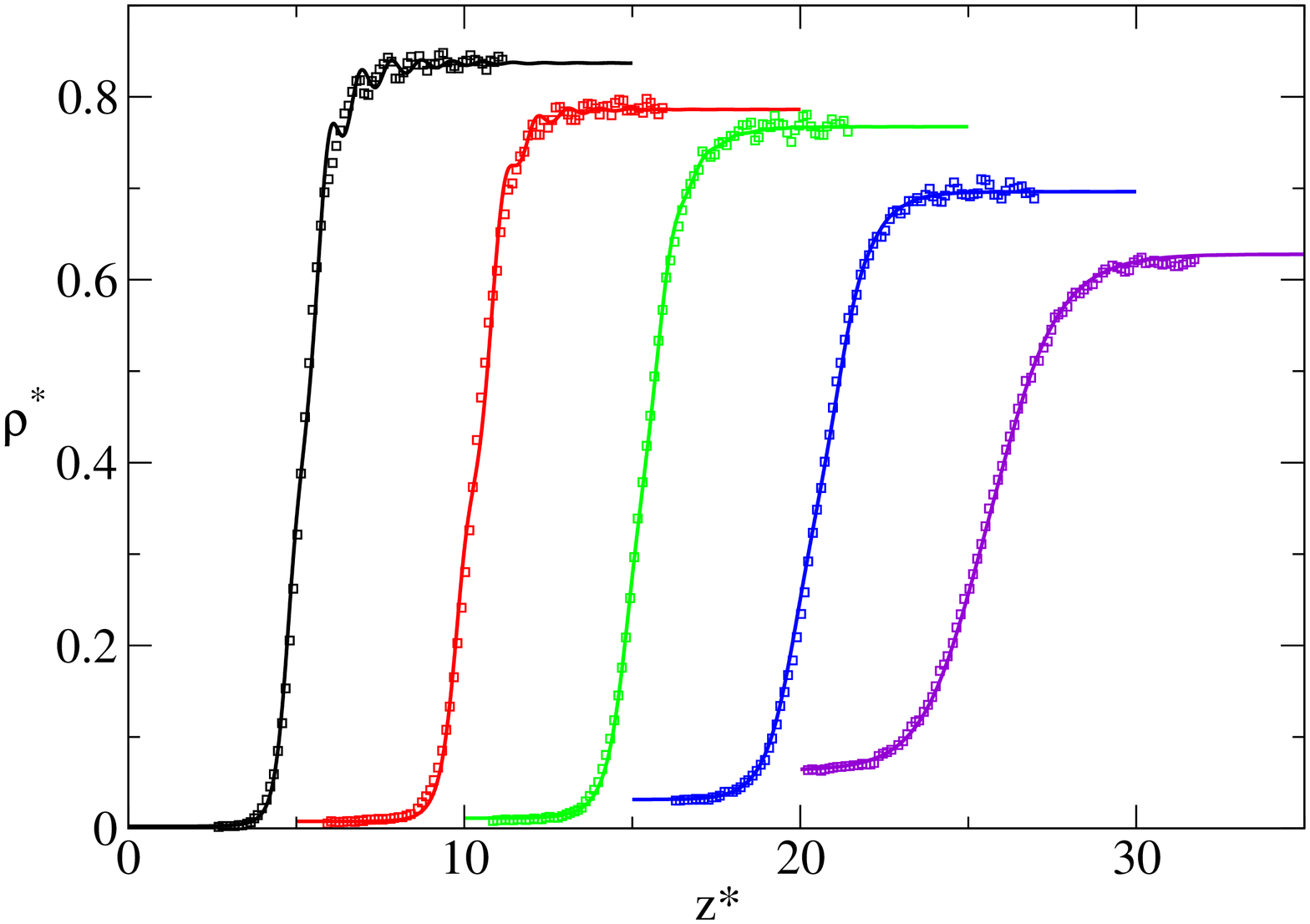}}}
\caption{}
\label{fig2a}
\end{figure*}

The density profile at the interface is shown in Fig.(\ref{fig2a}) for
several different temperatures and values of the cutoff. In all cases, the
theory is in good agreement with the profiles determined from simulation\cite%
{Mecke-LJ_Interface}. For $T^*=0.7$, near the triple point, and with a large
cutoff, the theory predicts oscillations in the profile at the interface, as
does for example the theory of Katsov and Weeks\cite{Katsov}. However, the
predicted oscillations here appear to be somewhat smaller than their
prediction and more in line with the profiles observed in simulation\cite%
{Mecke-LJ_Interface}. As the temperature is increased, the oscillations are
quickly suppressed. A similar effect results from using a shorter cutoff.

\subsection{Hard Wall}

The next test is the determination of the density profile for a fluid in
contact with a hard wall. In other studies, this comparison has been made
using the simulation data of Balabanic et al\cite{WallData}. However, as
this data is not readily available and as certain details such as the
potential cutoff are unclear\cite{TangScriven}, new simulations were
performed using the Grand Canonical Monte Carlo method\cite{FrenkelSmit}.
The simulation procedure consisted of several steps. First, the chemical
potential was estimated using the empirical equation of state. Then, an
initially random configuration of atoms was run for $10^{7}$ attempted Monte
Carlo moves with periodic boundary conditions. The simulation cell had
length $L$ in the $x$ and $y$ directions, and $nL$ in the $z$ direction
where $n$ is a parameter characterizing the geometry. This initial
equilibration was followed by a further equilibration of $10^{7}$ attempted
moves, but this time with hard walls at $z=0$ and $z=nL$. Finally, further
runs of $10^{7}$ attempted moves were performed during which the density
profile in the $z$ direction was tabulated after every $N$ attempted moves
using $200$ equally-spaced bins where $N$ is the expected average number of
atoms. To control for the effect of system size, runs were performed at all
densities involving approximately $2000$ atoms, corresponding to $n=2$, and
approximately $4000$ atoms corresponding to $n=4$. In all cases, the
potential cutoff was $r_{c}^*=4$, corresponding (roughly) to that used in
ref.\cite{WallData} (see the discussion in ref.\cite{TangScriven}).

Figures (\ref{wall-lo}) and (\ref{wall-hi}) show a comparison of the theory
(evaluated using the empirical equation of state) and simulation for
temperature $T^{\ast }=1.35$ and chemical potentials corresponding to bulk
densities of $\overline{\rho }^{*}=0.5,0.65$ and $0.82$ respectively. (These
conditions are the same as those used in previous studies\cite%
{WallData,TangScriven,Tang-LJ-DFT-2004}). The theory clearly captures the
very different qualitative behaviour across the range of densities and is
quantitatively accurate for the lower densities. At the highest density,
there is some difference in the calculated and measured profiles away from
the walls which can mostly be accounted for as an error in the phase of the
oscillations with the theory. This is interesting as the FMT for
hard-spheres gives, in the bulk, the Percus-Yevick structure and it has long
been known that the Percus-Yevick approximation for the the
pair-distribution funciton of bulk hard-spheres is also somewhat out of
phase at high densities\cite{HendGrundke}. Given that the MC-VDW is a
modified version of FMT, it is possible that the error in phase observed
here is related. It should also be noted that this phase error would be
partially or even totally negated if the points in the measured profile were
incorrectly plotted at the left coordinate of the bins rather than at there
center as has been done here. In all of the figures, it is clear that the
agreement between simulation and theory is particularly good near the wall
where the transition from drying at low bulk density to wetting at high
density is correctly predicted. This is not an accident as it is known that DFTs of this form satisfy
the exact sum
rule $\rho \left( 0\right) =\beta P_{bulk}$ where $P_{bulk}$ is the pressure
in the bulk far from the wall, a condition known as the wall theorem\cite%
{Swol,FundInhomLiq}.

\subsection{Slit Pore}

The final system considered is the Lennard-Jones fluid confined between two
infinite walls (i.e. a slit pore). Unlike the previous case of a hard wall,
the walls of the slit pore interact with the fluid via a modified
Lennard-Jones potential intended to mimic the interaction between the fluid
and a Lennard-Jones solid. The potential used here is the so-called 10-4-3
potential of Steele, 
\begin{equation*}
V_{wall}\left( z\right) =2\pi \varepsilon \left( \frac{2}{5}\left( \frac{%
\sigma }{z}\right) ^{10}-\left( \frac{\sigma }{z}\right) ^{4}-\frac{\sqrt{2}%
}{3\left( \frac{z}{\sigma }+0.61/\sqrt{2}\right) ^{3}}\right) ,
\end{equation*}%
which is specifically meant to model the interaction between the fluid and a
(100)\ plane in an FCC solid\cite{steele1,steele2,Snook1}. Since the
existing simulation data\cite{Snook1,Snook2,Magda} was obtained using very
small numbers of atoms, new GCMC simulations with much larger systems were
performed following the same protocol as for the hard wall. In the present
case, the intermolecular potential was cutoff at $r_{c}^*=6$ and no cutoff
was applied to the wall potential. Figures (\ref{slit-1}-\ref{slit-3}) show
a comparison between theory and simulation for the same conditions studied
in ref.\cite{Snook1,Snook2} and \cite{Magda}, namely the chemical potential
was set to the value corresponding to a bulk density of $\rho _{b}^{*}=0.5925
$ and the temperature was $T^* =1.2$. The slit sizes in the five
simulations are $H^*=3 , 4, 5, 6 $ and $10 $ and the simulations involved
approximately 1300,2100,2400, 3000 and 5000 atoms respectively.

The calculations are in good agreement with the
simulations. For the pores of intermediate size, the calculations tend to
overestimate the peak near the wall and to underestimate the amplitudes of
the subsequent oscillations but are nevertheless reasonable. These results
are in broad agreement with previous models such as that of Tang and Wu\cite%
{Tang-LJ-DFT-2004}. Further calculations, not shown here, confirm the
conclusions of Snook and van Megen\cite{Snook1} that the profiles are
insensitive to the value of the chemical potential so that errors in the
equation of state are not as important as in the case of the liquid-vapor
interface. Thus, the differences observed between theory and simulation must
be attributable to the theory itself and can serve as a sensitive test for
further improvements.

\bigskip

\section{Conclusions}

In this paper, the modified-core VDW approximation for the direct
correlation function has been used to construct a free energy functional
based on ideas from Fundamental Measure Theory. The resulting MC-VDW theory
was shown to give accurate predictions for the surface tension of the
liquid-vapor interface and the density profile near a wall and in slit pores.

There are several advantages to the MC-VDW free energy functional. It has
the practical advantage that it is no more complex than the mean field model
constructed using the FMT for the hard-sphere contribution. However, unlike
mean-field theory, it reproduces the input bulk free energy function so the
bulk thermodynamics are automatically correct. It also satisfies two
important exact relations. First, the exact relation between the free energy
functional and the direct correlation function of the bulk phase is
maintained: the second functional derivative of the free energy evaluated in
the bulk phase gives the correct bulk DCF. Second, the wall theorem -
relating the density at a hard wall to the pressure far from the wall - is
satisfied. These relations are difficult to preserve in theories which are
based on the introduction of the local density into bulk thermodynamic
relations such as those discussed in ref. \cite{Lutsko_JCP_2007}. The same
comment applies to theories which attempt to localize first order
perturbation theory as discussed, e.g., in ref.\cite{Wadewitz}. In fact, the
only theories based on these ideas which maintain the relation between the
free energy functional and the bulk DCF are those which eliminate all
density dependence beyond second order - which is basically the same as the
earliest perturbative DFT of Ramakrishnan and Yussouff\cite{RY}.

As discussed in ref. \cite{Lutsko_JCP_2007}, the modified-core DCF was
inspired by the work of Tang\cite%
{Tang-LJ_EOS,Tang-LJ-DFT-2003,Tang-LJ-DFT-2004,Tang-LJ-DFT-2005} on the
First Order Mean-Field Approximation (FMSA). In that approach, the
Ornstein-Zernicke equation is solved in a perturbative manner with the usual
mean-field closure conditions. This gives an analytic result for the DCF of
the bulk fluid which is similar to that used here, although the core
correction is not linear in the spatial variable. It has the advantage that
no input equation of state is needed, at least for Lennard-Jones
interactions. However, the mean-field liquid state theory is known to be
inaccurate in some circumstances\cite{HansenMcdonald} and the fact that good
results for the equation of state are obtained for Lennard-Jones by a first
order solution seems somewhat fortuitous. Furthermore, it still remains to
use the resulting bulk DCF to construct a DFT for inhomogeneous fluids.
Different theories seem to be required, e.g., for confined fluids\cite%
{Tang-LJ-DFT-2004} and for the liquid-vapor interface\cite{Tang-LJ-DFT-2005}%
. It is possible that the FMSA core correction could be used to construct a
MC-VDW theory similar to that studied here, although the more complex
analytic form of the core correction to the DCF will make this task more
difficult. Nevertheless, this would seem a promising approach.

The MC-VDW follows the philosophy described in the first paper in this series%
\cite{Lutsko_JCP_2007}: namely, that the bulk equation of state of fluids is
well-understood using, e.g., thermodynamic perturbation theory and that the
real goal of DFT should be the construction of functionals for inhomogeneous
systems. It would, of course, be more satisfying if, as in the case of the
FMSA cited above, the equation of state could be derived from the theory. It
is not impossible that this could be done using the present approach. Rather
than using the equation of state to determine the functions defining the
core correction (eqs.\ref{c1}-\ref{c3}), these might be determined by
imposing thermodynamic consistency between the resulting the free energy
functional and, say, the internal energy calculated using the
pair-distribution function given by the theory. This is the subject of
ongoing work.

A technical aspect of the MC-VDW model is that it is based on the simplest
form of FMT as originally proposed by Rosenfeld\cite{Rosenfeld1}. This
choice was made to avoid unneccessary complications in describing the model.
However, in future work such as in application to the solid phase, it might
prove necessary to use the more recent formulations of FMT\cite%
{Rosenfeld_1997_1,tarazona_2000_1,WhiteBear}, for the same reasons as occur
in the case of hard-spheres\cite{Rosenfeld_1997_1}.

In summary, the present approach has the virtue of satisfying the relevant
exact relations, of being conceptually and practically simple and of giving
quite reasonable results for a variety of model problems. In particular, a
single functional has been shown to be sufficient to describe both confined
fluids and the liquid-vapor interface.

\bigskip

\begin{acknowledgments}
This work was supported in part by the European Space Agency under contract
number ESA AO-2004-070.
\end{acknowledgments}

\bigskip \bigskip \appendix{}

\section{Derivation of model equations}

\label{Model}

Using the second of eq.(\ref{c1}) and eq.(\ref{c3}), one has that 
\begin{eqnarray}
j_{2}\left( \eta \right)  &=&-\frac{1}{2}\left( a_{0}+a_{1}\right) -3\left(
6\eta \right) j_{3}\left( \eta \right)  \\
j_{1}\left( \eta \right)  &=&\frac{1}{\overline{\rho }V}\beta
F_{core}^{x}+3\eta \left( a_{0}+a_{1}\right) +72\eta ^{2}j_{3}\left( \eta
\right)   \notag
\end{eqnarray}%
The third line of eq.(\ref{c1}) becomes%
\begin{eqnarray}
-a_{0} &=&j_{1}^{\prime }\left( \eta \right) +2\left( 6\eta \right)
j_{2}^{\prime }\left( \eta \right) +3\left( 6\eta \right) ^{2}j_{3}^{\prime
}\left( \eta \right)  \\
&=&\frac{d}{d\eta }\left( \frac{1}{\overline{\rho }V}\beta
F_{core}^{x}\right) -3\eta \left( a_{0}^{\prime }+a_{1}^{\prime }\right)
+3\left( a_{0}+a_{1}\right) -72\eta j_{3}\left( \eta \right) -36\eta ^{2}%
\frac{d}{d\eta }j_{3}\left( \eta \right)   \notag
\end{eqnarray}%
or, with some rearrangement,%
\begin{equation}
\frac{d}{d\eta }\left( 36\eta ^{2}j_{3}\left( \eta \right) -\frac{1}{%
\overline{\rho }}\beta f_{core}^{x}\right) =4a_{0}+3a_{1}-3\eta \left(
a_{0}^{\prime }+a_{1}^{\prime }\right) 
\end{equation}%
where $f_{core}^{x}=\frac{1}{V}\beta F_{core}^{x}$ is the contribution of
the core correction to the excess free energy density. 

For the chosen form of the tail-contribution to the DCF (which is
independent of density), the explicit expressions for $a_{0}$ and $a_{1}$ are%
\begin{eqnarray}
\frac{\pi d^{3}}{3}\left( 4a_{0}+3a_{1}\right)  &=&\frac{\partial
^{2}f_{HS}^{x}\left( \overline{\rho }\right) }{\partial \overline{\rho }^{2}}%
-\frac{\partial ^{2}f^{x}\left( \overline{\rho }\right) }{\partial \overline{%
\rho }^{2}}+4\pi \int_{d}^{\infty }\beta w\left( r\right) r^{2}dr \\
a_{0}+a_{1} &=&-\beta w\left( d\right)   \notag
\end{eqnarray}%
where $f^{x}\left( \overline{\rho }\right) =\frac{1}{V}\beta F^{x}$ is the
excess free energy density of the bulk fluid. Using these, the rhs can be
written as 
\begin{eqnarray}
&&4a_{0}+3a_{1}-3\eta \left( a_{0}^{\prime }+a_{1}^{\prime }\right)  \\
&=&\frac{3}{\pi d^{3}}\frac{\partial ^{2}\left( f_{HS}^{x}-f^{x}\right) }{%
\partial \overline{\rho }^{2}}+\frac{12}{d^{3}}\int_{d}^{\infty }\beta
w\left( r\right) r^{2}dr-3\eta \frac{d}{d\eta }\left( -\beta w\left(
d\right) -c_{HS}\left( d_{-};\overline{\rho };d\right) \right)   \notag
\end{eqnarray}%
Recognizing the contribution of the long-ranged tail to the bulk free energy
density,%
\begin{equation}
f_{tail}^{x}\equiv \frac{1}{2}\overline{\rho }^{2}\int w\left( r\right)
\Theta \left( r-d\right) d\mathbf{r,}
\end{equation}%
this can be written as%
\begin{equation}
4a_{0}+3a_{1}-3\eta \left( a_{0}^{\prime }+a_{1}^{\prime }\right) =\frac{3}{%
\pi d^{3}}\frac{\partial ^{2}\left( f_{HS}^{x}+f_{tail}^{x}-f^{x}\right) }{%
\partial \overline{\rho }^{2}}+3\eta \frac{d}{d\eta }c_{HS}\left( d_{-};%
\overline{\rho };d\right) 
\end{equation}%
or, since $f_{core}^{x}=f^{x}-f_{HS}^{x}-f_{tail}^{x}$,  
\begin{eqnarray}
\frac{d}{d\eta }\left( 36\eta ^{2}j_{3}\left( \eta \right) \right)  &=&\frac{%
d}{d\eta }\left( \frac{1}{\overline{\rho }}\beta f_{core}^{x}\right) -\frac{3%
}{\pi d^{3}}\frac{\partial ^{2}\beta f_{core}^{x}}{\partial \overline{\rho }%
^{2}}+3\eta \frac{d}{d\eta }c_{HS}\left( d_{-};\overline{\rho };d\right) 
\\
&=&\frac{d}{d\eta }\left[ \frac{1}{\overline{\rho }}\beta f_{core}^{x}-\frac{%
1}{2}\frac{\partial \beta f_{core}^{x}}{\partial \overline{\rho }}\right]
+3\eta \frac{d}{d\eta }c_{HS}\left( d_{-};\overline{\rho };d\right)  
\notag
\end{eqnarray}%
Finally, integrating gives%
\begin{equation}
36\eta ^{2}j_{3}\left( \eta \right) =\frac{1}{\overline{\rho }}\beta
f_{core}^{x}-\frac{1}{2}\frac{\partial \beta f_{core}^{x}}{\partial 
\overline{\rho }}-\left[ \frac{1}{\overline{\rho }}\beta f_{core}^{x}-\frac{1%
}{2}\frac{\partial \beta f_{core}^{x}}{\partial \overline{\rho }}\right] _{%
\overline{\rho }=0}+3\int_{0}^{\eta }\eta \frac{d}{d\eta }c_{HS}\left(
d_{-};\overline{\rho };d\right) d\eta 
\end{equation}%
where it is assumed that $j_{3}\left( \eta =0\right) =0$ so as to avoid
singulaties.

\bigskip 
\bibliographystyle{apsrev}
\bibliography{../physics}

\bigskip

\bigskip

\bigskip

\newpage
\section*{Figure captions}

Fig. 1. (Color on line) The coexistence curve for the Lennard-Jones fluid
as calculated using both the WCA perturbation theory, the BH theory and the
empirical JZG equation of state. The full lines are the liquid-vapor
coexistence curves, the dashed-lines are the spinodals and the symbols are
the simulation data from ref.\protect\cite{HansenLJPhaseDiagram}(circles)
and from ref. \protect\cite{Potoff_LJ_Interface}.

Fig. 2. (Color online)The surface tension as a function of temperature. The
symbols are measurements from simulations (circles from ref.\protect\cite%
{Duque-LJ-interface}, squares from ref.\protect\cite{Mecke-LJ_Interface},
diamonds from ref. \protect\cite{Potoff_LJ_Interface} and triangles from 
\protect\cite{HayeBruin}). The lines are the results of the MC-VDW model
evaluated with the JZG empirical equation of state (full line), the BH
perturbation theory (dashed line) and the WCA theory (dash-dotted line). The
lower curves and data are for a truncated and shifted potential with $%
r_c^*=2.5$

Fig. 3. (Color online) Density profiles at the liquid-vapor interface
calculated at different temperatures and values of the potential cutoff.
From left to right, the curves correspond to $T^*=0.7$ and $r_c^*=5.0$, $%
T^*=0.7$ and $r_c^*= 2.5$, $T^*=0.8$ and $r_c^*=5.0$, $T^*=0.8$ and $%
r_c^*=2.5$ and $T^*=1.1$ and $r_c^*=5.0$. The symbols are the data reported
in ref. \protect\cite{Mecke-LJ_Interface} and extracted from ref. 
\protect\cite{Katsov} as the original is no longer available\protect\cite%
{Fisher}. 

Fig. 4. (Color online) The structure of the fluid near a hard wall as
determined from simulation (symbols) and the theory (lines). The simulations
come from two runs each using cells with aspect ratio $1 \times 1 \times 2$,
circles, and $1 \times 1 \times 4$, squares. The upper curve and data are
for a chemical potential corresponding to bulk density $\protect\rho^*=%
\protect\rho\protect\sigma^3=0.65$ and the lower curve for density $\protect%
\rho^*=0.50$.

\begin{figure*}[h!tb]
\resizebox{12cm}{!}{\centerline{\includegraphics[angle=-90]{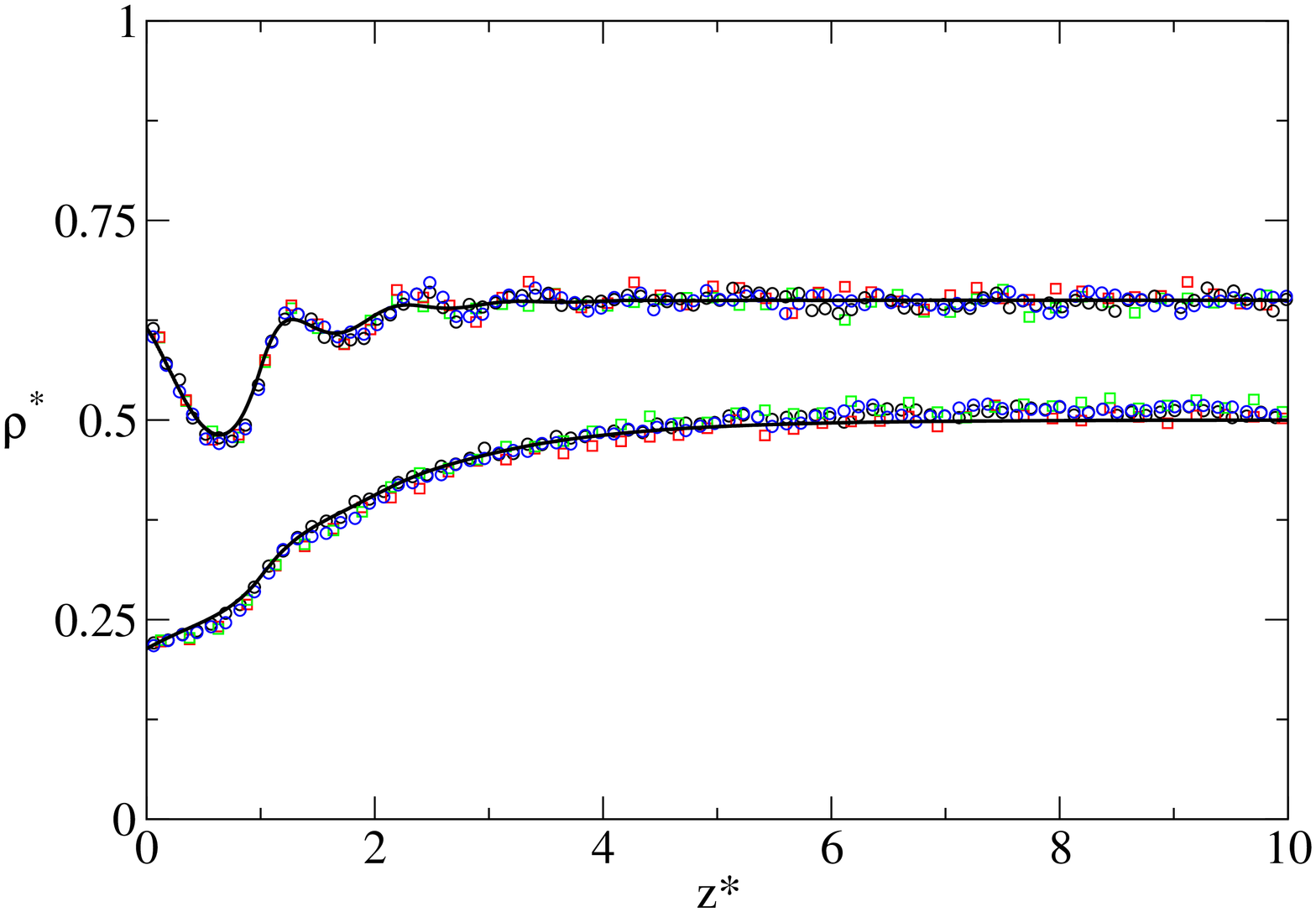}}}
\caption{}
\label{wall-lo}
\end{figure*}

\begin{figure*}[h!tb]
\resizebox{12cm}{!}{\centerline{\includegraphics[angle=-90]{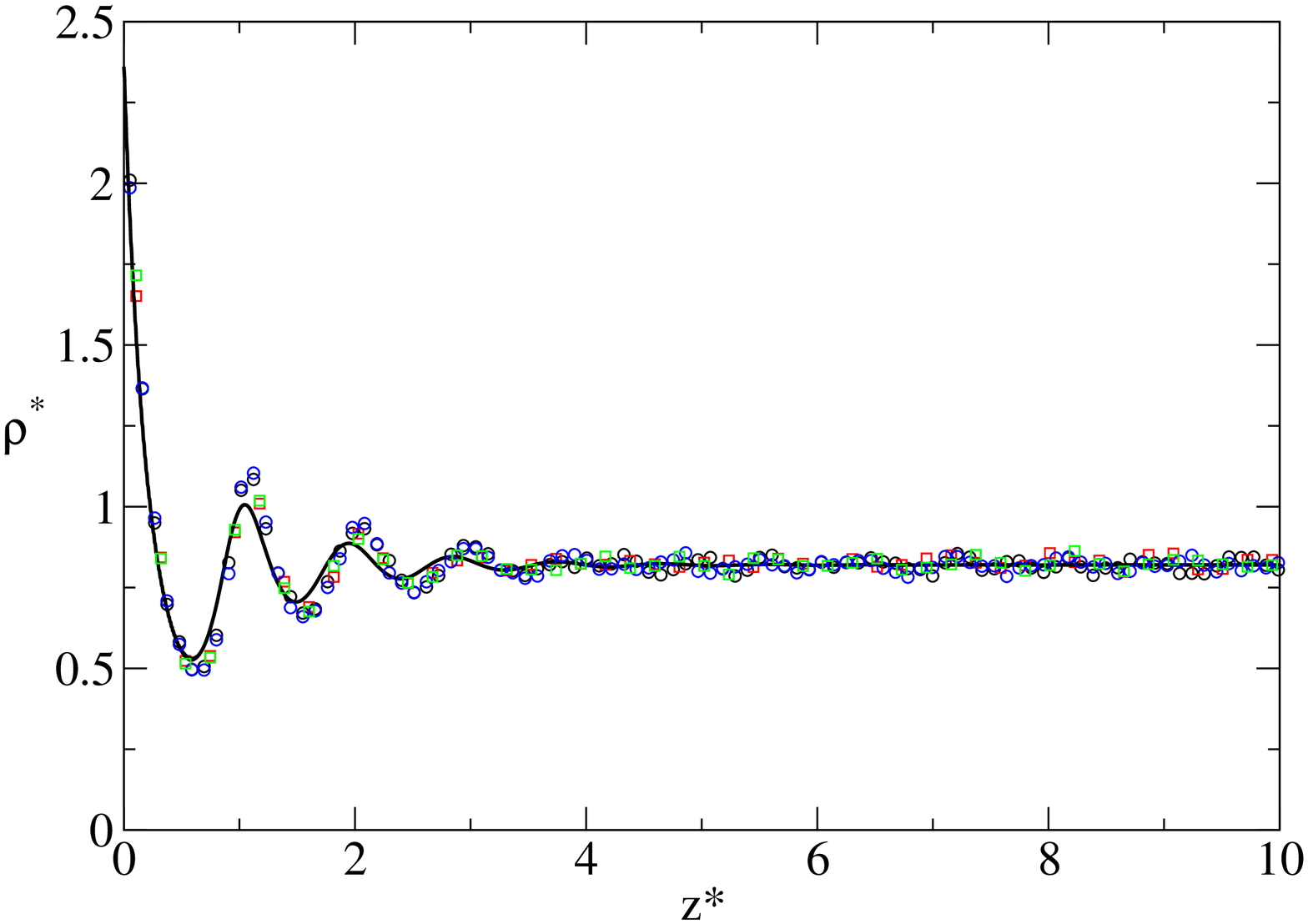}}}
\caption{}
\label{wall-hi}
\end{figure*}

Fig. 5. (Color online) Same as fig. (\protect\ref{wall-lo}) except that the
bulk density is $\protect\rho^*=0.85$.

Fig. 6. Comparison of the density distribution within slit pores of size $%
H^*=3$ and $H^*=4$ as calculated from the theory (lines) and as determined from
simulation (symbols).

\begin{figure*}[h!tb]
\resizebox{12cm}{!}{\centerline{\includegraphics[angle=-90]{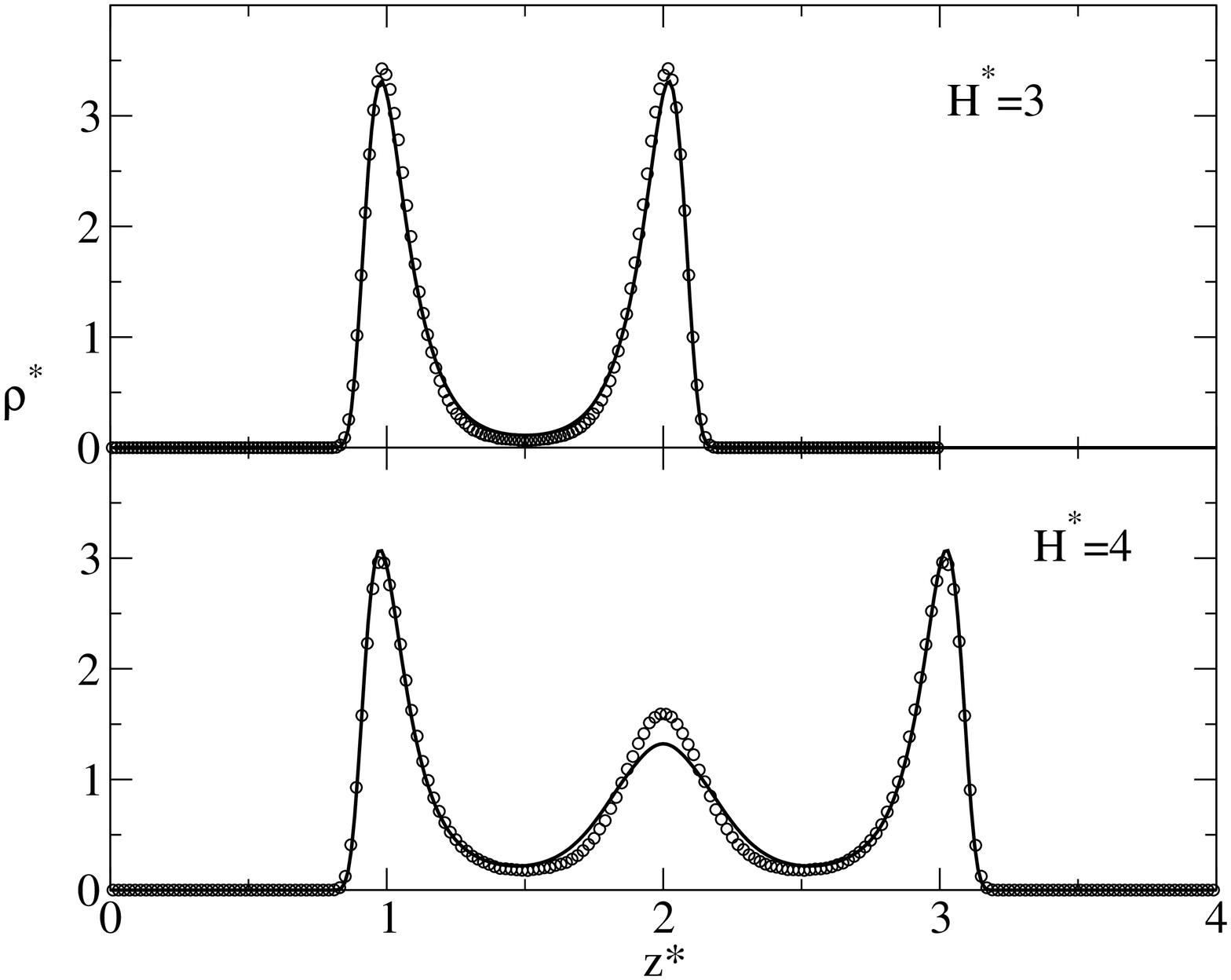}}}
\caption{}
\label{slit-1}
\end{figure*}

\begin{figure*}[h!tb]
\resizebox{12cm}{!}{\centerline{\includegraphics[angle=-90]{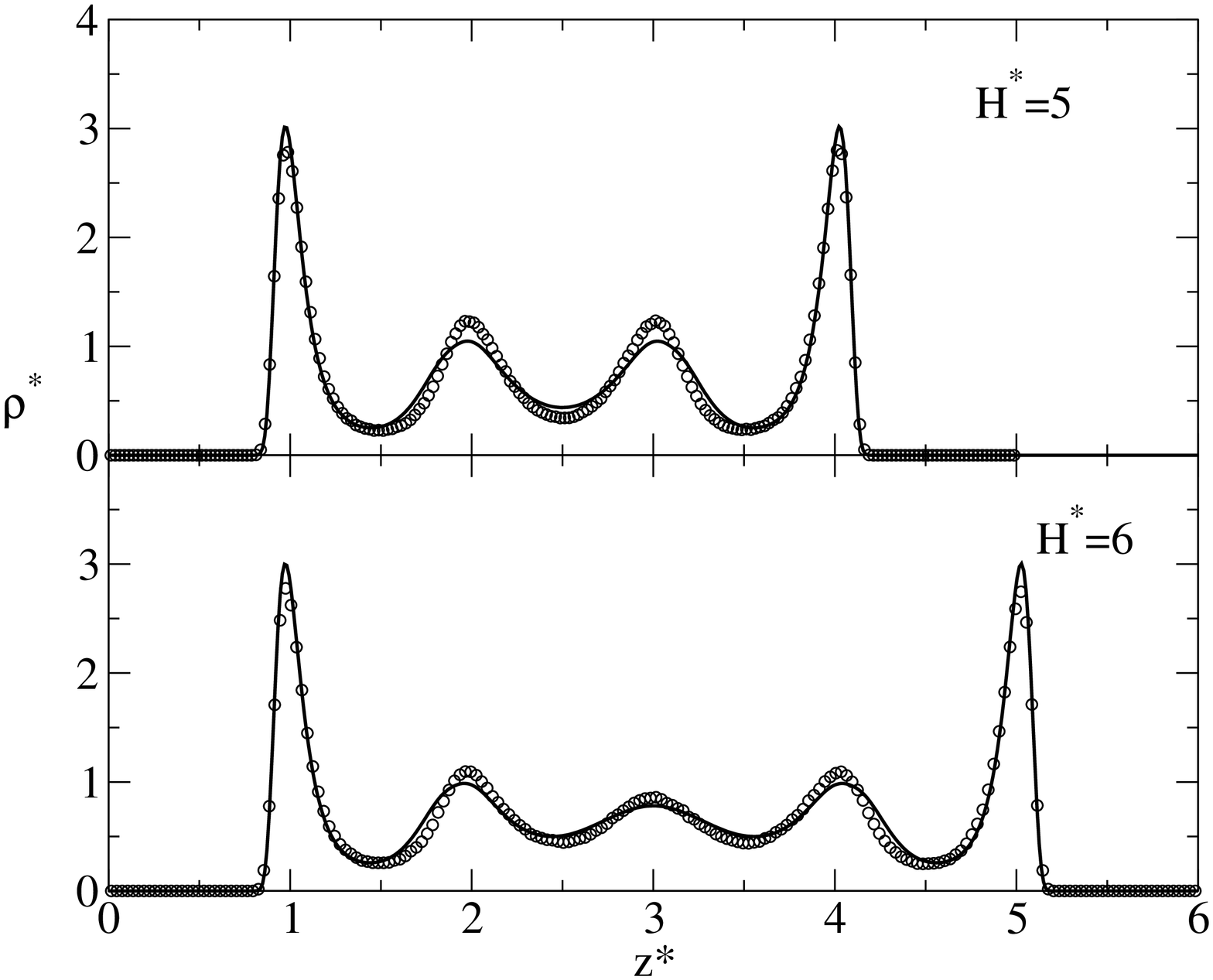}}}
\caption{}
\label{slit-2}
\end{figure*}

\begin{figure*}[h!tb]
\resizebox{12cm}{!}{\centerline{\includegraphics[angle=-90]{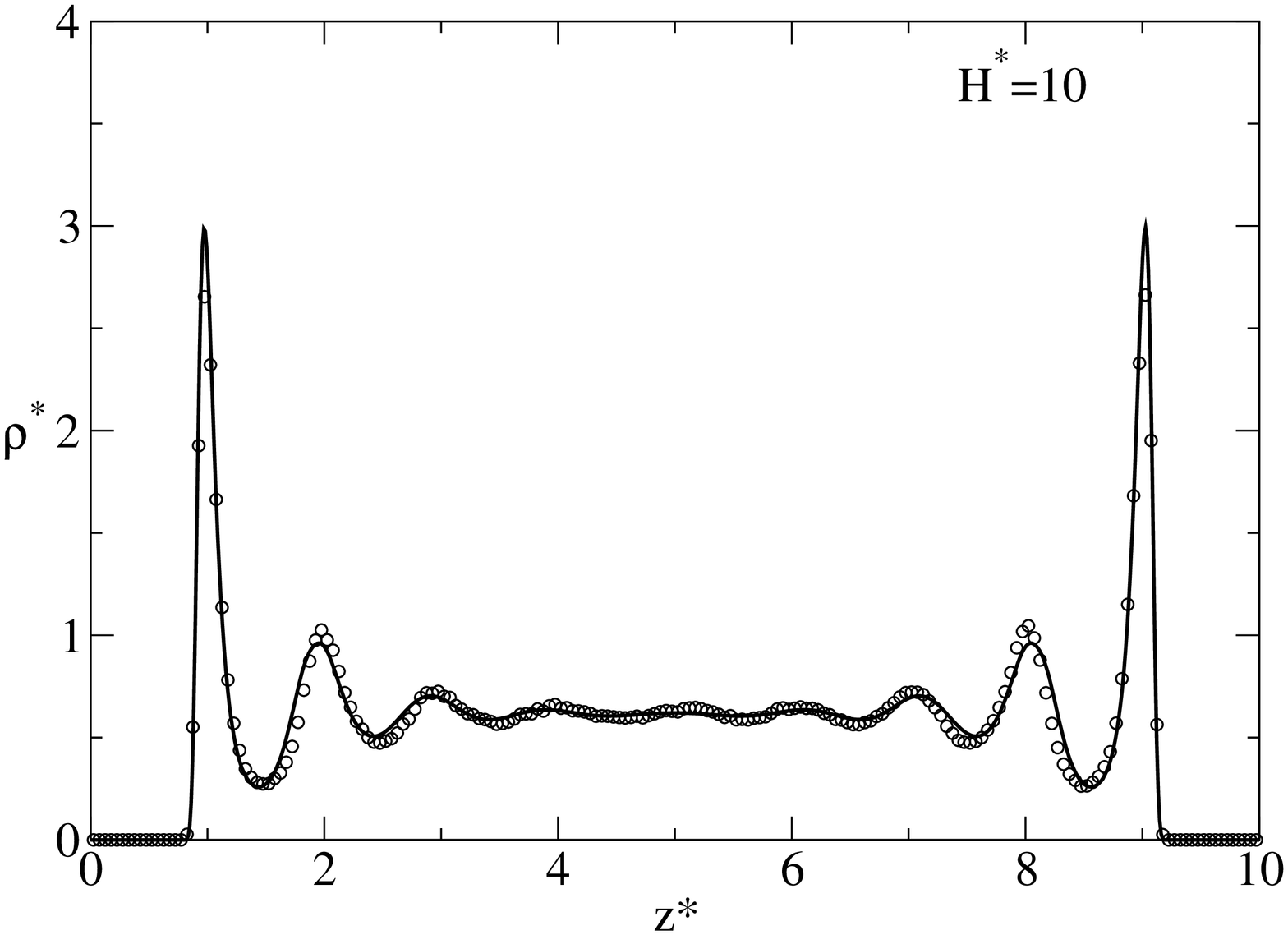}}}
\caption{}
\label{slit-3}
\end{figure*}

Fig. 7. Same as fig. (\protect\ref{slit-1}) for $H^*=5$ and $H^*=6$.

Fig. 8. Same as fig. (\protect\ref{slit-1}) for $H^*=10$.

\newpage

\end{document}